\documentclass{article}
\usepackage{amsfonts}
\usepackage{geometry}
\usepackage{indentfirst}
\usepackage{amsmath}
\usepackage{amssymb}
\usepackage[T1]{fontenc}
\usepackage{hyperref}
\usepackage[all]{xy}
\usepackage{color}

\usepackage{authblk}

\title{Remarks on simple interpolation between Jordanian twists}
\author[1]{Stjepan Meljanac\thanks{meljanac@irb.hr}}
\author[1]{Daniel Meljanac\thanks{Daniel.Meljanac@irb.hr}}
\author[2]{Anna Pacho\l \thanks{a.pachol@qmul.ac.uk}} 
\author[1]{Danijel Pikuti\'c\thanks{dpikutic@irb.hr}}
\affil[1]{Division of Theoretical Physics, Ru\dj{}er Bo\v{s}kovi\'c Institute, Bijeni\v{c}ka~c.54, HR-10002~Zagreb, Croatia}
\affil[2]{Queen Mary, University of London, School of Mathematical Sciences, Mile~End~Rd., London~E1~4NS, UK.}

\newcommand{\arxiv}[1]{\href{https://arxiv.org/abs/#1}{arXiv:#1}}

\begin{document}

\maketitle

\begin{abstract}
In this paper, we propose a simple generalization of the locally r-symmetric Jordanian twist, resulting in the one-parameter family of Jordanian twists. All the proposed twists differ by the coboundary twists and produce the same Jordanian deformation of the corresponding Lie algebra. They all provide the $\kappa$-Minkowski spacetime commutation relations. Constructions from noncommutative coordinates to the star product and coproduct, and from the star product to the coproduct and the twist are presented. The corresponding twist in the Hopf algebroid approach is given. Our results are presented symbolically by a diagram relating all of the possible constructions.
\end{abstract}

\section{Introduction}

In the Hopf algebras framework it is known that the given Hopf algebra $%
\mathcal{H}\left( \mu ,\Delta ,\epsilon ,S\right) $ can be deformed by using
the (Drinfeld) twist $\mathcal{F}\in \mathcal{H}\otimes \mathcal{H}$ \cite{drinfeld} which allows to deform
the coproduct and an antipode map in such a way that the compatibility
conditions are still satisfied and one gets a new Hopf algebra with deformed
maps. One special example of such twist is the so-called Jordanian twist $\mathcal{F}_{0}=\exp \left( \ln \left( 1+\alpha E\right) \otimes
H\right) $ with $\alpha $ as the deformation parameter \footnote{%
To distinguish different Jordanian twists we introduce a notation with sub-indices $0,1$, or $1/2$ which will become clear in section 2.}. It
was firstly constructed by O. Ogievetsky in \cite{Og} while studying the
moduli space of Hopf structures on the Borel subalgebra of $sl\left(
2\right) \ni \{H,E|\left[ H,E\right] =E\}$. This twist provides the simplest example of the triangular deformation of $sl\left(
2\right) $ algebra with the quantum R-matrix given by $R=\tilde{\mathcal F}\mathcal F^{-1}=1\otimes 1+r+\mathcal{O}%
\left( \alpha ^{2}\right) $, where $\mathcal{\tilde{F}}=\tau\circ \mathcal{F}$ is the flipped twist (with the flip map: $\tau :c\otimes d\rightarrow d\otimes c$) and $r$ is denoting the classical r-matrix.

%The  name Jordanian comes from the Jordanian quantum group [cite] which
%coacts on the noncommutative quantum plane .....

One can notice that $\mathcal{F}_{0}$ is not r-symmetric \cite{tolstoy} in the
sense that the term of the expansion in the deformation parameter $\alpha$, at the first order, is not given by the full
classical r-matrix $r$, i.e. is not of the form: $\mathcal{F}=1\otimes
1+\frac{1}{2}r+\mathcal{O}\left( \alpha ^{2}\right) $. However it can
be symmetrized by the so called coboundary twist $\mathcal{F}_{\omega
}=\omega ^{-1}\otimes \omega ^{-1}\Delta (\omega )$ as: 
\begin{equation*}
\mathcal{F}_{r}^{\left( \omega \right) }:=\omega ^{-1}\otimes \omega ^{-1}%
\mathcal{F}_{0}\Delta \left( \omega \right),
\end{equation*}%
where $\omega =\sqrt{\mu [ (1\otimes S)\mathcal F_0]}$.
The change of the twist by the coboundary twist $\mathcal F_\omega$ 
does not provide a new Hopf algebra. It provides a new presentation in the co-algebraic sector only, i.e. new form of the coproduct $\tilde\Delta=\mathcal F_\omega \Delta \mathcal F_\omega^{-1}$.

%$F_{i}^{\left( 1\right) }S\left( F_{i}^{\left( 2\right) }\right) $ with
%the short notation for the twist $\mathcal{F}=F_{i}^{\left( 1\right)
%}\otimes F_{i}^{\left( 2\right) }.$

One of the symmetrized versions (i.e. ``locally r-symmetric'') of the
Jordanian twist $\mathcal{\tilde{F}}=\tau\circ \mathcal{F}_0$ known in the literature is due to V.~N.~Tolstoy~\cite{tolstoy} which appeared in the context of studying quantum
deformations of relativistic symmetries.
% and the corresponding classification of classical r-matrices~\cite{zakrzewski}. Also the so-called extended Jordanian
%twists were studied in \cite{kulish}.
The locally r-symmetric version of the Jordanian twist calculated therein has the form: 
\begin{equation}  \label{tol_twist}
\mathcal{F}_{1/2}=e^{\frac{\alpha }{2}(HE\otimes 1+1\otimes HE)\,}e^{H\otimes\ln
\left( 1+\alpha E\right) }e^{{-\frac{\alpha }{2}}\left(HE{\otimes
1+H\otimes E+E\otimes H+1\otimes }HE\right) }
\end{equation}%
Here $\omega=\exp(\frac{1}{2}\alpha HE)$ was used to construct the coboundary twist.
There exists another version of symmetrized $\mathcal{F}_{0}$ and it is due
to Giaquinto and Zhang \cite{GZ}, but it will not be the object of our study
here.

The Jordanian twist investigated, firstly in \cite{Og} and then in \cite%
{tolstoy}, reappeared in the context of the so-called $\kappa $-Minkowski
noncommutative spacetime \cite{bp-2009, kmps, bu-kim}, which is an algebra
of coordinate functions equipped in the noncommutative star-product leading to the following commutation relations $[\hat{x}^{\mu
},\hat{x}^{\nu }]=\frac{i}{\kappa}(v^{\mu }\hat{x}^{\nu }-v^{\nu }\hat{x}^{\mu })$ where the deformation parameter is $\frac{1}{\kappa} $ and $v^{\mu }$ is the vector on Minkowski spacetime $\mathcal{M}_{1,n-1}$ in $n$-dimensions such that $v^{2}\in \{-1,0,1\}$.
%(i.e. space with the metric tensor with Lorentzian signature). 
In physical applications $\kappa $
is usually interpreted as the Planck mass or
Quantum Gravity scale. The natural quantum symmetry of this
noncommutative spacetime is the $\kappa $-Poincar\'e quantum group \cite{Luk1}
and it constitutes one of the examples of deformed relativistic spacetime
symmetries.
In this paper we are interested in the relation with the symmetry of the $\kappa $-Minkowski spacetime therefore we shall work with the generators of relativistic
symmetries: dilatation $D$ and momenta $p_{\alpha }$ (instead of the generators $H$ and $E$ of the $sl\left( 2\right) $ algebra) satisfying the same
commutation relation, i.e. $\left[p_{\alpha }, D\right] =p_{\alpha }$ \footnote{The correspondence with the generators of $sl(2)$ algebra is $H\rightarrow -D$ and $E\rightarrow p_\alpha$.}. The
dilatation generator $D$ is included in the minimal extension of the
relativistic spacetime symmetry, the so-called Poincar\'e-Weyl symmetry: $\{M_{\mu \nu },p_{\mu },D\}$ as well as in the conformal algebra $\{M_{\mu \nu },p_\mu,D,K_\mu \}$, therefore the Jordanian twists have the support in both of these algebras and can be used in their deformations \cite{bp-2009}, \cite{PRD-conformal}, respectively. However for the presentation of our results it is enough to consider the Lie algebra 
$g$ generated only by the dilatation and momenta operators satisfying $\left[
p_{\alpha }, D\right] =p_{\alpha }$ and $[p_\mu,p_\nu]=0$.
Recently Jordanian twists also have been considered in application to gravitational theory. For example, in \cite{ijgmmp2016}, the Jordanian twist was used
to construct the noncommutative differential calculus providing metrics as the solutions of non-vacuum Einstein equations, including cosmological constant or spatial curvature cases.
Additionally, it can be shown \cite{pv} that the star product, reproducing
the $\kappa $- Minkowski Lie algebra, obtained by suitably reducing the
so-called Wick--Voros star product, is in fact the star product from the
Jordanian twist $\mathcal{F}_{0}$. 
Jordanian deformations also have become popular in the context of
applications in AdS/CFT correspondence \cite{MatYosh}.

%%%%%%%%%%%%%%%%%%%%%%

In this paper we are interested in the generalization of the symmetrized
version of Jordanian twist $\mathcal{F}_{1/2}$ from \cite{tolstoy}. We
introduce the real parameter characterizing a whole family of
Jordanian twists interpolating between the original Jordanian twist $%
\mathcal{F}_{0}=\exp \left( -\ln \left( 1-a^{\alpha }p_{\alpha }\right)
\otimes D\right) $ and Jordanian twist $\mathcal F_1=\tau \circ 
\mathcal F_0|_{-a^\alpha}=$ $\exp \left(-D\otimes \ln \left( 1+a^{\alpha }p_{\alpha }\right) \right) $, where $a^{\alpha }=\frac{1}{\kappa}v^\alpha$  (in section 2).
They both lead to the $\kappa $-Minkowski spacetime commutation relations.
%The dilatation generator $D$ is included in the minimal extension of the relativistic spacetime symmetry, the so-called Poincar\'e-Weyl symmetry: $\{M_{\mu \nu}, p_\mu, D\}$ as well as in the conformal algebra $\{M_{\mu \nu }, p_\mu, D, K_\mu \}$, therefore the Jordanian twists have the support in both of these algebras and can be used in their deformations \cite{bp-2009}, \cite{PRD-conformal}, respectively. 
 The aim of this paper is to 
present different methods to obtain the same family of twists\footnote{up to the right ideal},
developed in the framework of noncommutative spacetimes.

In section 2, a generalization of the r-symmetric twist from \cite{tolstoy} is given. Corresponding deformed Hopf algebra symmetry, star products and differential realizations for noncommutative coordinates are presented. This family of Jordanian twists is also constructed as one exponential formula. 
However, in general, the twist might not always be known (e.g. in examples coming from deformation quantization framework).  

In section 3, we provide an example of a method used to construct the twist.  Starting from realizations of noncommutative coordinates, obtained in section 2, we construct the corresponding star product and coproduct. Also, from the star product and the coproduct, presented in section 2 (and coinciding with those in section 3), inverses of corresponding twists are obtained. They differ from the ones presented in section 2 by the right ideal, but indeed give the same deformed Hopf algebra, star product and realization of noncommutative coordinates. At the end of section 3, twists in Hopf algebroid approach, in terms of momenta and noncommutative coordinates, are given, showing that the used techniques appear within this more general
framework.
% providing a broader picture of the used techniques.
In section 4, concluding remarks are presented.

\subsection{Notation and formalism}

Here we will present the notation and recall some standard formulas related
to the twist formalism which will be necessary for the following sections of
the paper.

The Lie algebra $g$ generated by the dilatation and momenta operators is defined by
the commutation relations: $$\left[ p_{\mu },D\right] =p_{\mu },\quad\left[ p_{\mu
},p_{\nu }\right] =0.$$ The differential representation of the generators is the following: 
$D=x^{\alpha }\partial _{\alpha }=x\cdot\partial$ and $p_{\mu }=-i\partial _{\mu }$. We
will also use the short notation for the momenta $p_{\mu }$ contracted\footnote{Note that here we are using the relativistic notation: $a^{\nu }b_{\nu
}=a\cdot b$ (summation over $\nu $ index is assumed).} with
the vector $a^{\nu }$ as $A=ia^{\nu }\partial _{\nu }=-a^{\nu }p_{\nu }=-a\cdot p$. We do not specify signature of the metric, therefore the metric may be of any signature.

%$\kappa $ is the mass parameter of the order of Planck mass), and %where $x$ and $\partial$ satisfy

The generator $D$ can be also rewritten in Heisenberg realization as $%
D=ix^{\alpha }p_{\alpha }=ix\cdot p$, in terms of  Heisenberg algebra $H$, i.e.: 
\begin{equation}\label{heisenberg}
\lbrack x^{\mu },x^{\nu }]=0,\qquad \lbrack p _{\mu },x^{\nu
}]=-i\delta _{\mu }^{\nu },\qquad \lbrack p _{\mu },p _{\nu }]=0.
\end{equation}

The deformation of the Hopf algebra $U\left( g\right) \left( \mu ,\Delta
_{0},\epsilon ,S_{0}\right) $ of the universal enveloping algebra of $%
g=\{p_{\alpha },D:\left[ p_{\alpha },D\right] =p_{\alpha },\left[ p_{\mu
},p_{\nu }\right] =0\}$ given by the twist element $\mathcal{F}\in U\left(
g\right) \left[ \left[ \frac{1}{\kappa }\right] \right] \otimes U\left(
g\right) \left[ \left[ \frac{1}{\kappa }\right] \right] $ into $U^{\mathcal{%
F}}\left( g\right) \left( \mu ,\Delta ,\epsilon ,S\right) $ is provided by
the deformation of the coproduct and antipode maps as follows: $\Delta h=%
\mathcal{F}\Delta _{0}h\mathcal{F}^{-1}$, $S(h)
=[\mu \left( (1\otimes S)\mathcal{F}\right) ]S_0(h) [\mu \left(
(S\otimes 1)\mathcal{F}^{-1}\right) ]$, where $h\in g$.

The algebra of coordinates $\mathcal{A}$ with multiplication map $m:\mathcal{%
A}\otimes \mathcal{A}\rightarrow \mathcal{A}$, i.e. the spacetime algebra is
the Hopf module algebra with the action $U\left( g\right) \otimes \mathcal{A}%
\rightarrow \mathcal{A}$ of the Hopf algebra $U\left( g\right) $ on the
module algebra $\mathcal{A\ni }f,g$ such that $h\triangleright (m(f\otimes
g))=m\left[ \Delta h(\triangleright \otimes \triangleright )(f\otimes g)%
\right] $, where $h\in g$ and 
the action $\triangleright $ is defined by 
\begin{equation}  \label{action}
p_{\mu }\triangleright f(x)=-i\partial _{\mu }\triangleright f(x)=-i\frac{%
\partial f(x)}{\partial x^{\mu }}\text{ and }D\triangleright f(x)=x^{\alpha }%
\frac{\partial f(x)}{\partial x^{\alpha }}
\end{equation}%
and the module acts on itself as $x^{\mu }\triangleright f(x)=x^{\mu }f(x).$

The algebra of functions $\mathcal{A}$ becomes noncommutative during the
twist deformation once the usual multiplication is replaced by the
star-product between the functions 
%$\left( \mathcal{A},\mu _{\star }\right) $
i.e.: 
\begin{equation}
f\star g=m_{\star }\left( f\otimes g\right) =m[\mathcal{F}%
^{-1}(\triangleright \otimes \triangleright )(f\otimes g)]  \label{stp}
\end{equation}%
for $f,g\in \mathcal{A}$. The star product is associative (due to the fact
that the twist $\mathcal{F}$ satisfies cocycle condition). 

\section{Simple one-parameter family of Jordanian twists%Generalization of Tolstoy twist
}

%Tolstoy twist \cite{tolstoy} can be generalized with the following 
Let us introduce the following one-parameter family of twists $\mathcal{F}_u \in U(g)[[\frac1\kappa]]\otimes U(g)[[\frac1\kappa]]$: 
\begin{equation}
\mathcal{F}_{u}=\exp \left( -u(DA\otimes 1+1\otimes DA)\right) \exp \left(
-\ln (1+A)\otimes D\right) \exp \left( \Delta _{0}(uDA)\right) ,\quad u\in 
\mathbb{R},  \label{twist-family}
\end{equation}

The twists $\mathcal F_u$ can be obtained from $\mathcal F_0$ by the transformation with the coboundary twist with the element $\omega = e^{uDA}$. Note that the deformation parameter $1/\kappa$ is included in $A$ for the purpose of simplified notation. The twists $\mathcal{F}_{u}$ are Drinfeld twists \cite{drinfeld} $\forall u\in \mathbb{R}$, they satisfy normalization and cocycle condition\footnote{It is based on the fact that the change of the twist by the coboundary twist $\mathcal F_\omega$ influences the change of the coproduct by $\tilde\Delta=\mathcal F_\omega \Delta \mathcal F_\omega^{-1}$ which is isomorphic to the coproduct $\Delta$. Also adding the real valued parameter $u$ will not influence the cocyclicity.} and generalize the construction by V. N. Tolstoy~\cite{tolstoy}, (cf. \eqref{tol_twist}). The correspondence of this family of the Jordanian twists \eqref{twist-family} and the r-symmetric twist $\mathcal{F}_{1/2}$ is given by taking $u=\frac{1}{2}$ in \eqref{twist-family}. For $u=0$, twist \eqref{twist-family} simplifies to $\mathcal{F}_{0}=\exp \left( -\ln (1+A)\otimes D\right) $ and for $u=1$, simplifies to the twist $\mathcal{F}_{1}=\tau \circ \mathcal F_0\vert_{-a}=\tau \mathcal F_0\vert_{-a} \tau =\exp \left(-D\otimes \ln(1-A)\right) $.

The inverse of the above family of twists is
\begin{equation}
\mathcal{F}_{u}^{-1}=\exp \left( -\Delta _{0}(uDA)\right) \exp \left( \ln
(1+A)\otimes D\right) \exp \left( u(DA\otimes 1+1\otimes DA)\right) ,\quad
u\in \mathbb{R}.
\end{equation}

Now we can use the standard formulae (from section 1.1) for describing the
deformation of the Hopf algebra maps of $U^{\mathcal{F}}\left( g\right)
\left( \mu ,\Delta ,\epsilon ,S\right) $. The coproducts, star products and realizations in the rest of the paper depend on the parameter $u$, but for the sake of simplicity, the $u$-dependence will be omitted.

\textbf{Deformed Hopf algebra}

Coproducts $\Delta p_{\mu }$, corresponding to the above twist, for any
parameter $u\in \mathbb{R}$ are 
\begin{align}
\Delta p_{\mu } =\mathcal{F}_{u}\Delta _{0}p_{\mu }\mathcal{F}_{u}^{-1}&=
\frac{p_\mu \otimes (1-uA)+(1+(1-u)A)\otimes p_\mu}
{1\otimes 1+u(1-u)A\otimes A}  \label{delta-p} \\
\Delta D =\mathcal{F}_{u}\Delta _{0}D\mathcal{F}_{u}^{-1}&=
\left(D\otimes\frac1{1-uA}+\frac1{1+(1-u)A}\otimes D\right)
\left(1\otimes1 + u(1-u)A\otimes A \right) \label{delta-D}
%(D \otimes (1-uA)+(1+(1-u)A)\otimes D)\frac1{(1\otimes1-u\otimes A)(1\otimes1+(1-u)A\otimes1)}
%{1\otimes 1 - 1\otimes uA + (1-u)A\otimes 1+u(1-u)A\otimes A}
\end{align}%
The coproduct is coassociative. The corresponding antipode is given by 
\begin{align}\label{S-p}
S(p_{\mu }) &=\frac{-p_{\mu }}{1+(1-2u)A} \\
S(D) &=-D-(1-u)AD+uDA
-\frac{u(1-u)^2 A^2}{(1+(1-u)A)(1-uA)} \label{S-D}
\end{align}%
The antipode is antihomomorphism. The corresponding counit is trivial, i.e. $\epsilon (p_{\mu })=0,\epsilon (D)=0$ and $\epsilon (1)=1$.

Interesting special cases are $u=0$ (cf.\cite{pv}) and $u=1$  (cf.\cite{bp-2009}).
\begin{itemize}
\item For $u=0$ we get:
\begin{align}
%\mathcal F_0^{-1}& = %\sum_{m=0}^{\infty }\frac1{m!}A^{m}\otimes D^{\lf m} =
%e^{\ln(1+A)\otimes D}\\
\Delta p_\mu &=p_\mu\otimes 1 + (1+A)\otimes p_\mu \\
\Delta D &=D\otimes1+\frac1{1+A}\otimes D \\
S(p_\mu) &= \frac{-p_\mu}{1+A}\\
S(D)&=-(1+A)D
\end{align}
\item For $u=1$:
\begin{align}
%\mathcal F_1^{-1}& =%\sum_{m=0}^{\infty }\frac1{m!}(-1)^m D^{\lf m}\otimes A^{m}=
%e^{D\otimes \ln (1-A)} \\
\Delta p_\mu &=p_\mu\otimes(1-A) + 1\otimes p_\mu \\
\Delta D &=D\otimes\frac1{1-A}+1\otimes D \\
S(p_\mu) &= \frac{-p_\mu}{1-A}\\
S(D)&=-D(1-A)
\end{align}
\end{itemize}
%where 
%\begin{equation}
%D^{\lf m}=\prod_{k=0}^{m-1}(D-k),\quad m\geq 1\quad \text{and} \quad D^{\langle 0\rangle }=1.
%\end{equation}

\textbf{Star product}

The inverse of the twist $\mathcal{F}_{u}^{-1}$ also provides the star
product between the functions, as indicated in equation \eqref{stp} 
%$\left( \mathcal{A},\mu _{\star }\right) $.
This star product is associative (due to the fact that the twist $\mathcal{F}%
_{u}$ \eqref{twist-family} satisfies cocycle condition).

When we choose our functions to be exponential functions $e^{ik\cdot x}$ and $e^{iq\cdot x}$, then we define new function $\mathcal{D}_{\mu }(k,q)$: 
\begin{equation}
e^{ik\cdot x}\star e^{iq\cdot x}=m\left[ \mathcal{F}^{-1}(\triangleright
\otimes \triangleright )(e^{ik\cdot x}\otimes e^{iq\cdot x})\right] =e^{i%
\mathcal{D}_{\mu }(k,q)x^{\mu }},  \label{eikx-star-eiqx}
\end{equation}%
where $k,q\in \mathcal{M}_{1,n-1}$ - $n$-dimensional Minkowski spacetime.

One can calculate explicitly that in the case of twist $\mathcal{F}_{u}$ %
\eqref{twist-family} the function $\mathcal{D}_{\mu }(k,q)$ is given by 
\begin{equation}
\mathcal{D}_{\mu }(k,q)=\frac{k_{\mu }(1+u(a\cdot q))+(1-(1-u)(a\cdot k))q_\mu}
{1+u(1-u)(a\cdot k)(a\cdot q)}.  \label{mathcal-D}
\end{equation}%
Note that the function $\mathcal{D}_{\mu }(k,q)$ can be seen as rewriting
the coproduct $\Delta p_{\mu }$ without using the tensor product notation
(denoting left and right leg by $k$ and $q$ respectively). Therefore the
relation between the coproduct $\Delta p_{\mu }$ and the function $\mathcal{D%
}_{\mu }(k,q)$ is given by 
\begin{equation}
\Delta p_{\mu }=\mathcal{D}_{\mu }(p\otimes 1,1\otimes p),  \label{Delta-p-D}
\end{equation}%
hence $\Delta p_{\mu }$ uniquely determines $\mathcal{D}_{\mu }(k,q)$. In the case $u=0$, $\mathcal D_\mu(k,q)=k_\mu + (1-a\cdot k)q_\mu$, while in the case $u=1$, $\mathcal D_\mu(k,q)=k_\mu(1+a\cdot q) + q_\mu$. Deformed addition of momenta is given by $(k\oplus q)_\mu=\mathcal D_\mu(k,q)$.

\textbf{Coordinates}

Noncommutative coordinates $\hat x^\mu$, corresponding to the twist $\mathcal F_u$ \eqref{twist-family}, are given by 
\begin{equation}  \label{x-from-twist}
\begin{split}
\hat{x}^{\mu }=m\left[ \mathcal{F}_{u}^{-1}(\triangleright \otimes 1)(x_{\mu
}\otimes 1)\right] & =x^{\mu }(1-uA)+ia^{\mu }(1-u)D(1-uA) \\
& =(x^{\mu }+ia^{\mu }(1-u)D)(1-uA)
\end{split}%
\end{equation}
Alternatively, we notice that they can also be obtained from the coproducts \footnote{Later on we introduce \eqref{Fnormal} from which this relation follows naturaly.}
\begin{equation}
\begin{split}
\hat x^\mu %& =m\left[ \mathcal F_h^{-1}(\triangleright \otimes 1)(x_{\mu }\otimes 1)\right]  \\
& =x^\mu +ix^\alpha m\left[ (\Delta -\Delta _0)p _\alpha (\triangleright \otimes 1)(x^\mu \otimes 1)\right]  \\
& =(x^{\mu }+ia^{\mu }(1-u)D)(1-uA). \label{alt-x}
\end{split}%
\end{equation}%

The noncommutative coordinates $\hat x^\mu$ satisfy %\begin{align}
\begin{equation}
\begin{split}
\lbrack \hat{x}^{\mu },\hat{x}^{\nu }]& =i(a^{\mu }\hat{x}^{\nu }-a^{\nu }%
\hat{x}^{\mu }), \\
\lbrack p_{\mu },\hat{x}^{\nu }]& =(-i\delta _{\mu }^{\nu }+ia^{\nu
}(1-u)p_{\mu })(1-uA).
\end{split}
\end{equation}
In the case $u=0$, $\hat x^\mu=x^\mu+ia^\mu D$, while in the case $u=1$, $\hat x^\mu = x^\mu(1-A)$.

Note that we can define another set of the noncommutative coordinates $\hat{y}^{\mu }$ coming from the flipped version of \eqref{twist-family} as 
\begin{equation}\label{y-from-twist}
\begin{split}
\hat{y}^{\mu }& =m\left[ \tilde{\mathcal{F}}_{u}^{-1}(\triangleright \otimes
1)(x^{\mu }\otimes 1)\right]  \\
& =x^{\mu }+ix^{\alpha }m\left[ (\tilde{\Delta}-\Delta _{0})p_{\alpha
}(\triangleright \otimes 1)(x^{\mu }\otimes 1)\right]  \\
& =(x^{\mu }-ia^{\mu }uD)(1+(1-u)A),
\end{split}%
\end{equation}%
where $\tilde{\mathcal F}_u=\tau \circ \mathcal F_u=\tau\mathcal F_u\tau$ and 
$\tilde\Delta=\tau\circ\Delta = \tau\Delta\tau$. Generators $\hat{y}_{\mu }$ define a dual coordinate algebra in
the sense: 
\begin{equation}
\lbrack \hat{y}^{\mu },\hat{y}^{\nu }]=-i(a^{\mu }\hat{y}^{\nu }-a^{\nu }%
\hat{y}^{\mu })\quad \text{and}\quad \lbrack \hat{x}^{\mu },\hat{y}^{\nu }]=0
\end{equation}
i.e. they obey $\kappa$-Minkowski commutation relations with $a^\mu \to -a^\mu$. The commutation relation $[p_{\mu },\hat{y}^{\nu }]$ follows from the realization \eqref{y-from-twist}. In the case $u=0$, $\hat y^\mu=x^\mu(1+A)$, while in the case $u=1$, $\hat y^\mu = x^\mu - ia^\mu D$.

\textbf{One exponent formula for a family of Jordanian twists %
\eqref{twist-family}}

Above family of twists $\mathcal F_u$ given by \eqref{twist-family} can
be written as 
\begin{equation}
\mathcal{F}_{u}=\exp \left( (D\otimes uA)\theta -((1-u)A\otimes D)\tilde{%
\theta}|_{(-a)}\right) ,  \label{one_exp}
\end{equation}%
where 
\begin{equation}
\theta =\sum_{n=0}^{\infty }\sum_{\substack{ k,l=0  \\ k+l=n}}%
^{n}c_{k,l}A^{k}\otimes A^{l}\times 
\begin{cases}
(1-u)^{k}u^{l} & \text{for }u\in \mathbb{R}\setminus\{0,1\} \\ 
\delta _{l0} & \text{for }u=0 \\ 
\delta _{k0} & \text{for }u=1%
\end{cases}%
\end{equation}

\begin{equation}
\tilde{\theta}=\tau (\theta) =\tau\circ\theta\circ\tau,
\end{equation}%
where $c_{k,l}\in\mathbb R$ and $c_{0,0}=1$.
The first three terms in the above
expansion in $\frac{1}{\kappa }$ %, determining the coefficients $c_{k,l}$ in $\theta $,
 are: 
\begin{align}
\left(\ln\mathcal F_u\right)_1 &=D\otimes uA-(1-u)A\otimes D, \\
\left(\ln\mathcal F_u\right)_2 &=\frac{1}{2}\left[ 
%D\otimes u^2 A^2+(D\otimes 1+1\otimes D)[(1-u)A\otimes uA]+(1-u)^{2}A^{2}\otimes D
D\otimes uA + (1-u)A\otimes D
\right]
\left[
1\otimes uA + (1-u)A\otimes 1
\right],
\\
\begin{split}
\left(\ln\mathcal F_u\right)_3 &=D\otimes uA\left[ \frac13
(1\otimes u^2 A^2)+\frac16(1-u)A\otimes uA-\frac{1}{6}(1-u)^{2}A^{2}\otimes 1\right]  \\
& -(1-u)A\otimes D\left[ \frac13(1-u)^2 A^2\otimes 1+\frac16
(1-u)A\otimes uA-\frac16(1\otimes u^2 A^2)\right] .
\end{split}
\end{align}
In special cases $u=0$ and $u=1$, $\mathcal F_u$ reduces to
\begin{equation}
\mathcal F_0 = e^{-\ln(1+A)\otimes D}, \qquad \mathcal F_1 = e^{-D\otimes\ln(1-A)},
\end{equation}
respectively.

The corresponding quantum R-matrix is $\mathcal{R}=\tilde{\mathcal{F}}_{u}\mathcal{F}
_{u}^{-1}=1\otimes 1+r+\mathcal{O}\left( \frac{1}{\kappa ^{2}}\right) $, where $r=A\otimes D-D\otimes A$, $\forall u\in \mathbb{R}$.

\section{From realization to star product and twist}

Realizations of noncommutative coordinates $\hat x^\mu$ can be generally expressed in terms of Heisenberg algebra $H$ \eqref{heisenberg}, generated by $x^\mu$ and $p_\mu$. If $\hat x^\mu$ generate a Lie algebra, there exists universal formula for $\hat x^\mu$, related to Weyl ordering~\cite{durov}.\footnote{For a more general case, see \cite{mercati}.} Starting from the realization \eqref{x-from-twist} (see for example \cite{stojic})
\begin{equation}\label{hat-x-s3}
\hat{x}^{\mu }=x^{\alpha }\varphi _{\alpha }{}^{\mu }(p)=(x^{\mu }+ia^{\mu
}(1-u)D)(1-uA),
\end{equation}
one can reconstruct the star product using the following method. 
Recalling the action introduced in section 2 (\ref{action}) we can explain how
the exponent function of the noncommutative coordinates acts on the usual
exponential function, via the realization (\ref{x-from-twist}) of $\hat{x}$. 
\begin{equation}
e^{ik\cdot \hat{x}}\triangleright e^{iq\cdot x}{=e^{ik\cdot x^{\alpha
}\varphi _{\alpha }{}^{\mu }(p)}\triangleright e^{iq\cdot x}}=e^{i\mathcal{P}%
(k,q)\cdot x},
\end{equation}%
Here we introduced another set of functions, denoted by $\mathcal{P}_{\mu
}(k,q)$ which satisfy the following differential equations \cite{mmss, svrtan, EPJC2015, mercati}
\begin{equation}
\frac{d\mathcal{P}_{\mu }(\lambda k,q)}{d\lambda }=\varphi _{\mu }{}^{\alpha
}(\mathcal{P}(\lambda k,q))k_{\alpha }.
\end{equation}
Note that the equation involves the same function $\varphi_\mu{}^\alpha(p)$ from the realization of the noncommutative coordinates. The boundary conditions are 
$\mathcal{P}_{\mu }(0,q)=q_{\mu }$. The solution is 
\begin{equation}
\mathcal{P}_{\mu }(k,q)=\frac{K_{\mu }(k)(1+u(a\cdot q))+(1-(1-u)(a\cdot K(k)))q_{\mu }}
{1+u(1-u)(a\cdot K(k))(a\cdot q)},
\end{equation}%
where 
\begin{equation}
K_{\mu }(k)=\mathcal{P}_{\mu }(k,0)=k_{\mu }\frac{e^{a\cdot k}-1}{a\cdot k}%
\frac{1}{(1-u)e^{a\cdot k}+u}.
\end{equation}%
The inverse function of $K_{\mu }(k)$, defined as $K_\mu(K^{-1}(k))=K^{-1}_\mu(K(k))=k_\mu$, is given by 
\begin{equation}
K_{\mu }^{-1}(k)=k_{\mu }\frac{1}{a\cdot k}\ln \left( \frac{1+u(a\cdot k)}{%
1-(1-u)a\cdot k}\right) .
\end{equation}%
There exists a relation between the function $\mathcal{P}_{\mu }(k,q)$ and
introduced before function $\mathcal{D}_{\mu }(k,q)$ as $\mathcal{D}(k,q)=%
\mathcal{P}(K^{-1}(k),q)$. Thanks to this, we can rewrite the corresponding
star product of exponential functions as 
\begin{equation}
e^{ik\cdot x}\star e^{iq\cdot x}=e^{iK^{-1}(k)\cdot \hat{x}}\triangleright
e^{iq\cdot x}=e^{i\mathcal{P}(K^{-1}(k),q)\cdot x}=e^{i\mathcal{D}(k,q)\cdot
x}.  \label{star-D}
\end{equation}
The function $\mathcal{D}_{\mu }(k,q)$ from equation \eqref{star-D}
coincides with equation \eqref{mathcal-D}.

From the function $\mathcal{D}_{\mu }(k,q)$, the coproduct $\Delta p_{\mu }$
%is obtained via $\Delta p_{\mu }=\mathcal{D}_{\mu }(p\otimes 1,1\otimes p)$ 
can be recovered using \eqref{Delta-p-D}
and antipodes $S(p_{\mu })$ follow analogously. Alternatively, it is possible to find the coproduct $\Delta p_\mu$ as%we find 
\begin{equation}\label{adx}
\Delta p_{\mu }=e^{iK_{\alpha }^{-1}(p)\otimes \mathrm{ad}_{\hat{x}^{\alpha
}}}(1\otimes p_{\mu }),
\end{equation}

{\bf From star product and coproduct to twist}

From equations \eqref{Delta-p-D} and \eqref{star-D} it follows \cite{govindarajan, IJMPA2014, EPJC2015}:
\begin{equation}\begin{split}\label{star-lim}
e^{ik\cdot x}\star e^{iq\cdot x}
&= \lim_{\substack{y\to x\\z\to x}}\left(
e^{
ix\cdot\left[
\mathcal D(-i\partial^y,-i\partial^z)+\partial^y+\partial^z 
\right]
}
(e^{ik\cdot y}e^{iq\cdot z})
\right) \\
&=m\left[
:e^{i((1-t)\otimes x^\alpha+tx^\alpha\otimes1)(\Delta-\Delta_0)p_\alpha}:
(\triangleright\otimes\triangleright)
( e^{ik\cdot x}\otimes e^{iq\cdot x})
\right], \quad \forall t\in\mathbb R .
\end{split}\end{equation}
Note that on the right hand side of the above
equation the terms multiplied by $t$ cancel each other due to the normal
ordering and the multiplication map $m$.

From this identity and equations \eqref{stp}, \eqref{eikx-star-eiqx}, it follows:
\begin{equation}\label{Fnormal}
\mathcal F^{-1}=
:e^{i((1-t)\otimes x^\alpha+tx^\alpha\otimes1)(\Delta-\Delta_0)p_\alpha}:
+\mathcal I_0,
\end{equation}
where $\mathcal I_0\subset H\otimes H$ is the right ideal of the coordinate algebra $\mathcal A$, defined by
\begin{equation}
m\left[ \mathcal I_0 (\triangleright\otimes\triangleright)(\mathcal A\otimes \mathcal A) \right] = 0
\end{equation}

For the case $u=0$, with $t=0$, twist $\mathcal F_0^{-1}$ is given by
\begin{equation}
\mathcal F_0^{-1} = :e^{A\otimes D}: =e^{\ln(1+A)\otimes D},
\end{equation}
while for the case $u=1$, with $t=1$, twist $\mathcal F_1^{-1}$ is given by
\begin{equation}
\mathcal F_1^{-1} = :e^{-D\otimes A}: =e^{D\otimes \ln(1-A)}.
\end{equation}
Calculation of $\mathcal F^{-1}$ in equation \eqref{Fnormal} in the form $\mathcal F^{-1}= e^{-f}$ for linear realizations of $\hat x^\mu$ is presented in \cite{EPJC2015} and for Abelian twists in \cite{govindarajan}.

{\bf Twist in the Hopf algebroid approach}

Deformed phase spaces of Lie algebra type are presented and studied in \cite{mss,lukierski, Lu, Bohm}. Twists in Hopf algebroid approach were studied and constructed in \cite{IJMPA2014, skoda-mel, rina-PLA2013, Xu}. 

Generally, in Hopf algebroid approach, the twist is given by 
\begin{equation}\label{Falgebroid}
\mathcal{F}^{-1}= e^{-ip_\alpha \otimes x^\alpha}
e^{iK_\gamma^{-1}(p)\otimes \hat x^\beta} + \mathcal I_0 = e^{-ip_\alpha \otimes x^\alpha}
e^{iK_\gamma^{-1}(p)\otimes x^\beta \varphi_\beta{}^\gamma(p)} + \mathcal I_0
\end{equation}

This is a generalization of the result presented in \cite{skoda-mel}. The full Hopf algebroid analysis will be presented elsewhere.

\section{Concluding remarks}

Inspired by the recent interest in Jordanian deformations, in this paper we focused on the simple generalization of the locally r-symmetric Jordanian twist. By introduction of one real valued parameter $u$ we obtained the family of Jordanian twists which provides interpolation between the original Jordanian twist (for $u=0$) and its flipped version (for $u=1$, up to minus sign in the deformation parameter). All of the proposed twists provide the $\kappa$-Minkowski spacetime and have the support in the Poincar\'e-Weyl or conformal algebras as deformed symmetries of this noncommutative spacetime. Another important issue considered here was the presentation of different methods relating the twist, deformed coproducts with the star-product and the realizations for the noncommutative coordinates.

It is important to note that we can present our results symbolically by
the following diagram: %\begin{displaymath}
%\begin{equation}
%\begin{split}
%\xymatrix{ 
%\hat x^\mu = x^\alpha \varphi_\alpha{}^\mu(p) \ar[r] \ar[rd]\ar[d] & \Delta p_\mu \ar[d] \ar[ld] \ar[l] \\ 
%e^{ik\cdot x}\star e^{iq\cdot x} \ar[u] \ar[ru] \ar[r] & \mathcal F \ar[l] \ar[lu] \ar[u]}  \notag
%\end{split}
%\end{equation}
%\end{displaymath}
\begin{equation}
\begin{split}
\xymatrix{ 
&\!\!\!\!\! \!\!\!\!\!\hat x^\mu =  x^\alpha \varphi_\alpha{}^\mu(p) \ar[ldd] \ar[rdd] \ar[d] \!\!\!\!\! \!\!\!\!\!&\\ 
&\mathcal F \ar[u] \ar[ld] \ar[rd]&\\
e^{ik\cdot x}\star e^{iq\cdot x} \ar[ruu] \ar[rr] \ar[ru]&&\Delta p_\mu \ar[luu] \ar[ll] \ar[lu]}  \notag
\end{split}%
\end{equation}
%If we start with the expression for $\Delta p_\mu$ \eqref{delta-p}, equivalent to star product $e^{ik\cdot x}\star e^{iq\cdot x}=e^{i\mathcal{D}(k,q)\cdot x}$ \eqref{eikx-star-eiqx}, results for twist $\mathcal{F}^{-1}$ and realizations $\hat x^\mu$... % are the same as in the section 3. 
In section 2, starting from the twist \eqref{twist-family}, we have found deformed Hopf algebra symmetry \eqref{delta-p}-\eqref{S-D}, star products \eqref{eikx-star-eiqx}, \eqref{mathcal-D} and corresponding realizations for noncommutative coordinates \eqref{x-from-twist}. Relation between coproduct $\Delta p_\mu$ and star products $e^{ik\cdot x}\star e^{iq\cdot x}$ \eqref{eikx-star-eiqx}, \eqref{mathcal-D} are given in \eqref{Delta-p-D}. Noncommutative coordinates $\hat x^\mu$ are also obtained from the coproduct $\Delta p_\mu$ in equation \eqref{alt-x} and also from the related star product $e^{ik\cdot x}\star e^{iq\cdot x}=e^{i\mathcal D(k,q)\cdot x}$. In section 3, starting from realizetions of noncommutative coordinates \eqref{hat-x-s3}, we have constructed corresponding star products \eqref{mathcal-D}, \eqref{star-D} and coproducts. Alternatively, coproduct $\Delta p_\mu$ is obtained from noncommutative coordinates \eqref{adx}. Also, from the star product and the coproduct in section 2, see also equations \eqref{star-D}, \eqref{star-lim}, inverses of the corresponding twist \eqref{Fnormal} are obtained. Twists in the Hopf algebroid approach, in terms of momenta and noncommutative coordinates, are given in equation \eqref{Falgebroid}. Note that the inverse twist $\mathcal F^{-1}$ is not uniquely determined from the star product, coproduct $\Delta p_\mu$ and noncommutative coordinates $\hat x^\mu$ and it is determined up to a right ideal $\mathcal I_0$. %____

More general realizations are of the type $\hat x^\mu = x^\alpha \varphi_\alpha{}^\mu(p)
+ \chi^\mu(p)$, but we restricted our considerations to the case $%
\chi^\mu(p)=0$ and the above diagram corresponds to $\chi^\mu(p)=0$. We
point out that all our results are exact. There are few possible extensions
of these results, leading to new insights of the deformation quantization.
However, they require an extension to the Hopf algebroid framework.  Drinfeld twists can be lifted to Hopf algebroids \cite{bp_jpa2016} from the Hopf algebra, but it is worth to note that not all the twists obtained from the coproduct are the Hopf algebra twists. The issues of the twist reconstruction within Hopf algebroids and the generalization of the other version
of symmetrized Jordanian twist presented in \cite{GZ} will be thoroughly discussed in another paper.

%Comments about Hopf algebroid, twists...

\section*{Acknowledgements}

Authors would like to thank A.~Borowiec for useful comments. 
The work by S.M. and D.P. has been supported by the Croatian Science Foundation under the Project No. IP-2014-09-9582 as well as by the H2020 Twinning project No 692194, ``RBI-T-WINNING''.
A.~P. acknowledges the
funding from the European Union's Horizon 2020 research and innovation
programme under the Marie Sklodowska-Curie grant agreement No 660061.

\end{document}